\documentclass{iau_FM}
\usepackage{graphicx}

\newcommand{\ca}{\mbox{Ca\,{\sc ii}~K\,}}

\title[Polar field of the Sun] 
{Estimation of  Polar Magnetic Fields using Ca II K Polar Network as a Proxy}

\author[Dibya et al.]   
{Dibya Kirti Mishra$^{1,2}$, Bibhuti Kumar Jha$^3$, Theodosios Chatzistergos$^4$, Ilaria Ermolli$^5$, Dipankar Banerjee$^{1,6,7}$
 \and M. Saleem Khan$^2$}

\affiliation{$^1$Aryabhatta Research Institute of Observational Sciences, Nainital-263002, Uttarakhand, India \\ email: {\tt dibyakirtimishra@aries.res.in} \\[\affilskip]
$^2$Mahatma Jyotiba Phule Rohilkhand University, Bareilly-243006, Uttar Pradesh, India 
$^3$Southwest Research Institute, Boulder, CO 80302, USA\\
$^4$Max Planck Institute for Solar System Research, Justus-von-Liebig-Weg 3, D-37077 Göttingen,Germany\\
$^5$INAF Osservatorio Astronomico di Roma, Via Frascati 33, 00078 Monte Porzio Catone, Italy\\
$^6$Indian Institute of Astrophysics, Koramangala, Bangalore 560034, India\\
$^7$Center of Excellence in Space Sciences India, IISER Kolkata, Mohanpur 741246, West Bengal, India
}

\pubyear{2024}
\jname{Astronomy in Focus Issue A32,Focus Meeting 8} 
\editors{Alexander Kosovichev, ed.}
\begin{document}

\maketitle

\begin{abstract}
The polar magnetic field plays a crucial role in the solar dynamo model and contributes to predicting future solar cycles. However, continuous and direct measurements of this polar field have been available only since 1976, with data provided by the Wilcox Solar Observatory (WSO). Recent findings suggest that the \ca\ Polar Network Index (PNI) can serve as a promising proxy for estimating the polar field of the Sun. In this study, we aim to reconstruct the polar field for the pre-1976 period by leveraging \ca\ data from the Kodaikanal Solar Observatory (KoSO; 1904\,--\,2007) and modern \ca\ observations from the Rome Precision Solar Photometric Telescope (Rome-PSPT; 2000\,--\,2022). We employ an automatic adaptive threshold technique to detect polar networks and calculate PNI values. Then, we calibrate these PNI values with the WSO polar field to reconstruct the polar field over 119 years.

\end{abstract}

\firstsection 
\section{Introduction}

The polar field plays a crucial role in the solar dynamo model, as it subsequently generates the toroidal field, leading to sunspot formation (\cite[Charbonneau 2020]{Charbonneau2020}). The strength of the polar field at solar minimum is strongly correlated with the sunspot maximum in the next cycle, aiding in predictions of future solar cycle strengths (\cite[Kumar \etal\ 2021]{kumar2021}, \cite[Kumar \etal\ 2022]{kumar2022}, \cite[Upton \& Hathaway 2023]{upton2023}). Recently, \cite[Jha \& Upton (2024)]{jha2024} used polar field data to predict the strength of Solar Cycle\,25, which further shows the significance of the polar field.

Accurately measuring the polar field is challenging due to the Sun-Earth line observation constraint. Despite this constraint, direct and continuous measurements started in 1976 with the Wilcox Solar Observatory (WSO; \cite[Svalgaard \etal\ 1978]{Svalgaard1978}). But, for periods before 1976, researchers have to rely on various proxies, such as photospheric polar faculae and polar filaments etc., with polar faculae being the most commonly used proxy for polar field estimation based on Mount Wilson Observatory (MWO) white-light data (1906\,--\,2007, \cite[Muñoz-Jaramillo \etal\ 2013]{munoz2013}). However, this approach has limitations, as manual counting methods introduce inconsistencies in faculae counts for different observatories. Consequently, developing a more robust proxy for improved polar field estimation is essential.

Recently, the polar network index (PNI) has been identified as a superior proxy for polar field estimation compared to polar faculae due to its automated identification process (\cite[Priyal \etal\ 2014a]{priyal2014a}). In this study, they used \ca\ data from the Kodaikanal Solar Observatory (KoSO; 1909\,--\,1990) for PNI estimation. However, certain periods in the KoSO dataset (\cite[Priyal \etal\ 2014a]{priyal2014a}) contained incorrect timestamps, potentially compromising the accuracy of PNI and polar field estimation in the polar regions. Therefore, to make a more accurate polar field reconstruction, we utilize recently calibrated (\cite[Chatzistergos \etal\ 2018]{Chatzistergos2018}, \cite[Chatzistergos \etal\ 2019]{Chatzistergos2019}, \cite[Chatzistergos \etal\ 2019b]{Chatzistergos2019b}, \cite[Chatzistergos \etal\ 2020]{Chatzistergos2020}) and rotation-corrected (\cite[Jha 2022]{jha2022}, \cite[Jha \etal\ 2024]{Jha2024}) KoSO \ca\ data (1904\,--\,2007) in this study.

\begin{figure}[htb!]
\includegraphics[scale=0.5]{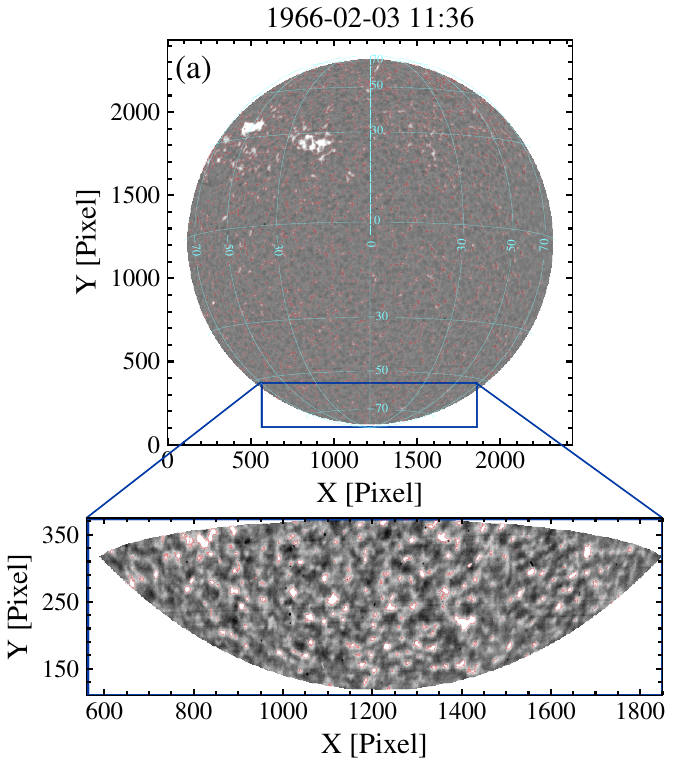}
\label{fig1}
\includegraphics[scale=0.85]{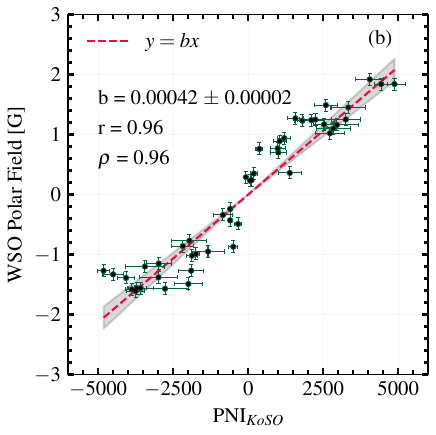}
\label{fig2}
\caption{(a) The red contours represent detected features, including plages and network regions, in the \ca\ image from KoSO, observed on 1966-02-03 11:36 IST. The rectangular blue box indicates the selected region covering latitudes from -55 degrees to -90 degrees in the southern pole. The red contours highlight the identified bright polar network regions within this designated area. (b) The scatter plots illustrate the relationship between PNI from KoSO and the polar field measurements from WSO from 1977 to 2000. The red line represents the best-fit line, while the shaded areas denote the 95\% confidence interval.}
\end{figure}

\section{Data and Methodology}

The KoSO holds an extensive archive of \ca\ spectroheliograms with a bandpass of 0.05\,nm centered at 393.367\,nm (\cite[Chatterjee \etal\ 2016]{Chatterjee2016}, \cite[Jha 2022]{jha2022}, \cite[Mishra \etal\ 2024]{Mishra2024}). For this study, we use these century-long \ca\ data, which have recently been calibrated using a new technique (\cite[Chatzistergos \etal\ 2020]{Chatzistergos2020}) and corrected for solar rotation (\cite[Jha \etal\ 2024]{Jha2024}. A sample full-disk image is presented in Fig.\,\ref{fig1}(a). Additionally, we incorporate \ca\ observations from the Rome Precision Solar Photometric Telescope (Rome-PSPT) for the period 2000\,--\,2022, with a bandpass of 0.25\,nm centered at 393.37\,nm, covering data from 1996 to the present (\cite[Ermolli \etal\ 2022]{Ermolli2022}).

We employ an adaptive thresholding technique to identify polar network bright regions, specifically the NR method introduced by \cite[Nesme-Ribes \etal\ (1996)]{Nesme-Ribes1996}. First, we take the rotation-corrected image and crop the solar disk within 0.98\,$R$ to exclude unwanted regions near the limb. We then apply Eq.\,\ref{eq1} to apply the threshold and detect bright network regions, where ($\overline{C}_{\mathrm{min}}$) represents the minimum mean contrast, and ($\sigma_{\mathrm{min}}$), the minimum standard deviation, characterizes the background intensity of quiet Sun. Here, $m_{\mathrm{n}}$ is set to 3.2 for both KoSO and Rome-PSPT images. The identified bright network regions are shown in Fig.\,\ref{fig2}(b). Further details on the method are available in \cite[Mishra \etal\ (2024)]{mishra2024}.

\begin{equation}
T=\overline{C}_{\mathrm{min}} + m_{\mathrm{n}}*\sigma_{\mathrm{min}},
\label{eq1}
\end{equation}

We analyze the latitude range of $\pm (55^{\circ}-90^{\circ})$ for both the northern and southern poles, selecting data during August–September for the north and February–March for the south. A sample of polar network detection in the southern pole, outlined in red contours, is shown in Fig.\,\ref{fig1}(a). For calculating the PNI within the chosen region, we use Eq.\,\ref{eq2}, where $R$ represents the solar disk radius in pixels. Using this method, we compute the PNI for both the KoSO (1904\,--\,2007) and Rome-PSPT (2000\,--\,2022) datasets.

\begin{equation}
{\rm PNI} = \frac{{\rm Number~of~Network~Pixels}}{\pi R^2}\times10^6,
\label{eq2}
\end{equation}

\section{Result}

\begin{figure}[htb!]
\centering
\includegraphics[scale=0.88]{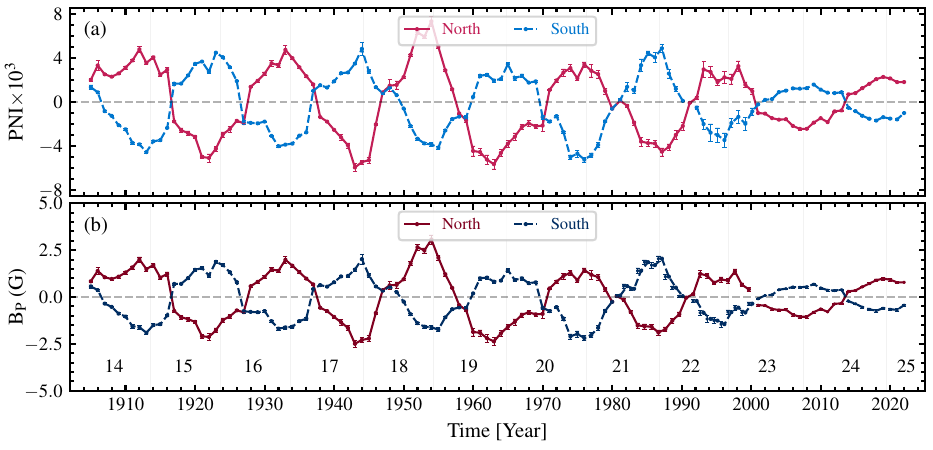}
\caption{(a) The polarity-signed PNI series derived from \ca\ data from KoSO and Rome-PSPT is presented for both the northern and southern hemispheres. (b) The polar magnetic fields reconstructed from the PNI of KoSO and Rome-PSPT are illustrated for the time period from 1904 to 2022, with the northern hemisphere (dark red) and the southern hemisphere (dark blue).}
\label{fig3}
\end{figure}

A two-month averaged PNI series is created by combining \ca\ data from KoSO and Rome-PSPT spanning 1904 to 2022. Since the PNI values are all positive, we determine polarity based on the timing of the minimum PNI values in both hemispheres. The resulting composite polarity-signed PNI series, covering the period from 1904 to 2022, is presented in Fig.\,\ref{fig3}(a) for both hemispheres. To reconstruct the polar field from the PNI values, which is the primary objective of this study, we calibrate the PNI values against the direct polar field measurements from WSO. A calibration plot for the KoSO PNI values is shown in Fig.\,\ref{fig2}(b) with green points, demonstrating a strong correlation between the PNI and polar field values, with Pearson ($r$) and Spearman ($\rho$) correlation coefficients of 0.96. We apply a linear fitting to this relationship, yielding calibration constant values of $b=0.00042\pm0.00002$, as illustrated in Fig.\,\ref{fig2}(b). This methodology allows us to reconstruct the polar fields using PNI data from 1904 to 2022 for both hemispheres, as depicted in Fig.\,\ref{fig3}(b).

\section{Conclusion}

In this study, we utilized \ca\ data from KoSO (1904\,--\,2007) and Rome-PSPT (2000\,--\,2022) to identify bright polar network regions within the latitude range of $\pm (55^{\circ}-90^{\circ})$ for both the northern and southern poles. We calculated PNI from the detected polar network regions and constructed a polarity-signed PNI series. Ultimately, we reconstructed the polar fields using the composite PNI values over approximately 11 solar cycles, beginning in 1904. This reconstructed polar field information will enhance our understanding of the solar dynamo model, which has previously been limited by a lack of historical polar field data.

\section{Acknowledgment}

We extend our gratitude to the IAU for waiving the registration fee, which enabled me to attend the General Assembly in person in Cape Town, South Africa.

\end{document}